\documentclass[ reprint,amsmath,amssymb,aps,prr, superscriptaddress,]{revtex4-2} 

\usepackage{bbold}
\usepackage{amsmath}
\usepackage{amssymb}
\usepackage{lipsum}
\usepackage[cjk]{kotex}
\usepackage{graphicx}
\usepackage{dcolumn}
\usepackage{xcolor}
\usepackage{bm}
\usepackage[mathscr]{euscript}
\usepackage{comment}
\usepackage{tikz}
\usetikzlibrary{quotes,angles}
\usepackage{hyperref}
\usepackage{ulem}
\usepackage{youngtab}
\newcommand{\vacuum}{\left | \emptyset \right >}

\begin{document}

\title{Exact hole-induced  $SU(N)$ flavor-singlets in certain $U=\infty$ $SU(N)$ Hubbard models
}

\author{Kyung-Su Kim (김경수)}
\affiliation{Department of Physics, Stanford University, Stanford, California 93405, USA}
\author{Hosho Katsura (\begin{CJK*}{UTF8}{ipxm}桂法称\end{CJK*})}
\affiliation{Department of Physics, Graduate School of Science, The University of Tokyo, 7-3-1 Hongo, Tokyo 113-0033, Japan}
\affiliation{Institute for Physics of Intelligence, The University of Tokyo, 7-3-1 Hongo, Tokyo 113-0033, Japan}
\affiliation{Trans-scale Quantum Science Institute, The University of Tokyo, 7-3-1, Hongo, Tokyo 113-0033, Japan}

\date{\today}

\begin{abstract}

We prove that the motion of a single hole induces $SU(N)$ flavor singlets in the $U=\infty$ $SU(N)$ (Fermi) Hubbard model on a Husimi-like tree graph.
The result is generalized to certain $t$-$J$ models with singlet hopping terms typically neglected in the literature.
This is an $SU(N)$ generalization of the ``counter-Nagaoka theorem" introduced in [Phys. Rev. B \textbf{107}, L140401 (2023)]. 
Our results suggest the existence of resonating flavor singlet (RFS)-like polarons in the $t$-$J$ models on a more realistic non-bipartite lattice.
Such RFS polarons may be relevant for a novel strong-coupling mechanism of superconductivity or other exotic fractionalized phases of matter.
\end{abstract}
\maketitle

\section{Introduction} \label{sec:intro}

The $SU(2)$ Hubbard model in the presence of a hole doping has been  extensively studied
 as it is expected to capture essential features of the high-temperature superconductivity in cuprate superconductors \cite{ anderson1987RVB, leeNagaosaWen, keimerKivelsonNorman, arovas2021hubbard}.
Despite its deceptively simple form, the model presents significant challenges and complexity due to competing tendencies to develop various types of distinct ordered phases  \cite{fradkinKivelsonTranquada}.
Even in the strong coupling ($U=\infty$) limit, an analytical solution on a bipartite lattice (e.g., square lattice) exists only for single-hole doping on a finite-sized system---the celebrated ``Nagaoka theorem" states that such a system leads to a fully polarized ferromagnet \cite{nagaoka1966ferromagnetism, tasaki1989extension, fazekas1999lecture, tasaki2020physics, von2010probing, dehollain2020nagaoka}. 
In a physical context, the Nagaoka theorem implies the formation of the ferromagnetic Nagaoka polaron, which has been observed in numerics \cite{liu2012phases, white2001Npolaron} and in cold-atom experiments \cite{koepsell2019NpolaronColdAtom}.

On the other hand, it is known that the hole motion in the $U=\infty$ $SU(2)$ Hubbard model on a {\it non-bipartite} lattice (e.g., triangular lattice) induces antiferromagnetic correlations around it \cite{takano1989RVB, haerter2005kineticAF, sposetti2014kineticAF, zhu2022doped}.
However, for such a non-bipartite lattice, even the single-hole problem is poorly understood due to the frustration inherent in antiferromagnetism.
The problem has been recently solved in a frustration-free version of a non-bipartite lattice, which unambiguously demonstrated that a hole is surrounded by resonating valence bond (RVB)-like correlations  \cite{kim2023RVB}. 
Such a result suggests the formation of an RVB polaron on a more realistic non-bipartite lattice.

\begin{figure}[b]
    \centering
\includegraphics[width = 0.48 \textwidth]{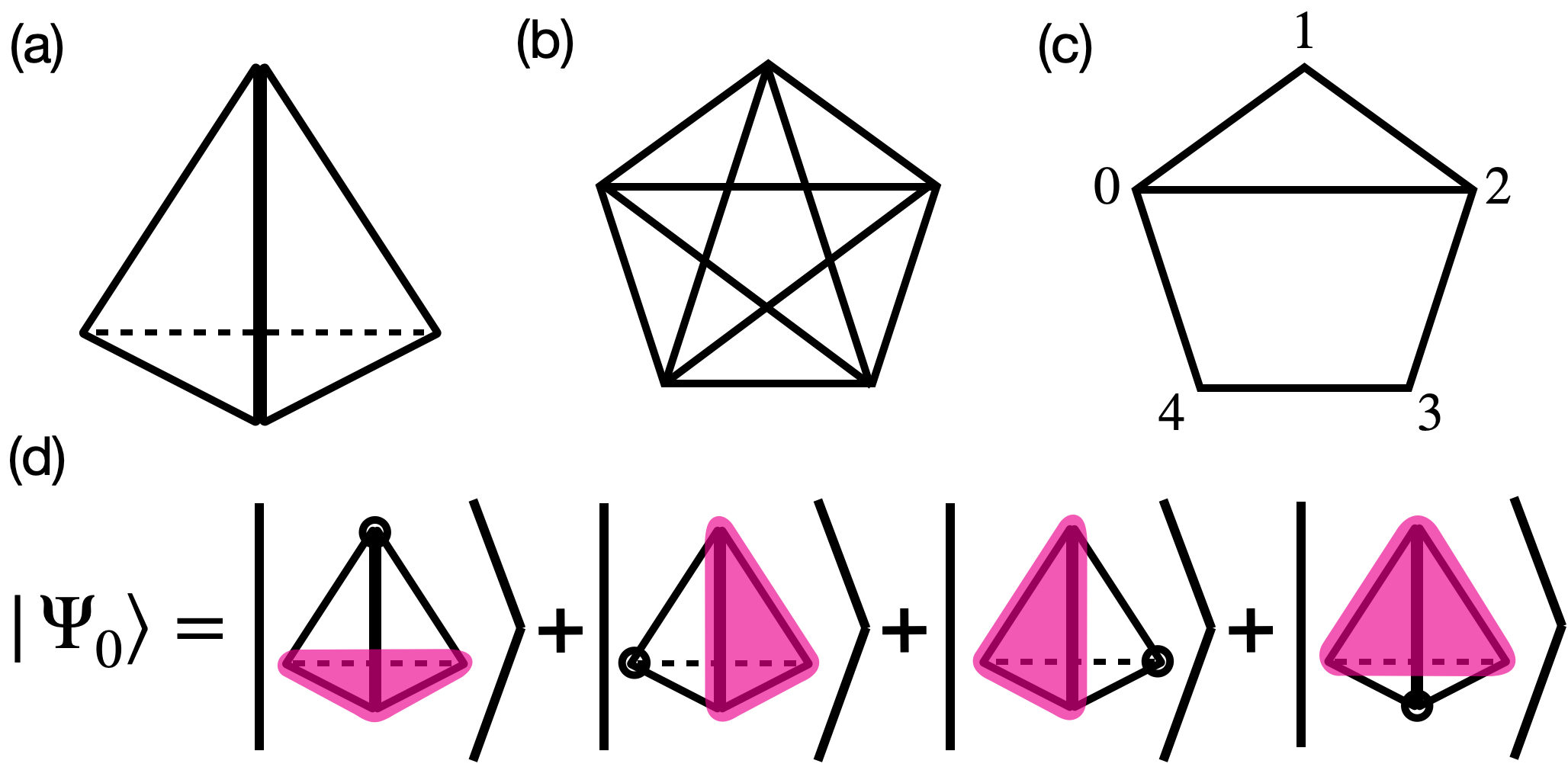}
    \caption{(a),(b) Examples of a complete graph with fully connected edges. (c) An example of a non-complete graph which nevertheless satisfies the connectivity condition. (d) The ground state of the $U=\infty$ $SU(3)$ Hubbard model in the presence of a single hole on a tetrahedron with uniform $t_{ij}= t$ and $\hat V =0$ in Eq. (\ref{eq:Hamiltonian}). Magenta trimers denote $SU(3)$ flavor-singlets, i.e., three fermions with complete flavor-antisymmetry, and circles denote the location of the hole. The signs associated with the many-body states appearing in $\left |\Psi_0 \right >$ are defined implicitly in Eq. (\ref{eq:flavor-singlet}).  
    }
    \label{fig:1}
\end{figure}

For systems with an emergent (or exact) $SU(N)$ symmetry with $N>2$ \cite{gorshkov2010SUN,taie20126,scazza2014SUN, zhang2014SUN, hofrichter2016direct, ozawa2018antiferromagnetic}, e.g., systems with degenerate multiple valleys or flavors \cite{calvera2022WC, kangVafek2019,vafekKang2020,bultinck2020ground, bernevig2021twisted}, their physics may be characterized by the $SU(N)$ Hubbard model or its generalizations under suitable circumstances.
If so, the magnetism at the $\frac 1 N$th filling (one fermion per site)  in the strong coupling regime, $U\gg t$, is captured by the $SU(N)$ Heisenberg model with exchange interactions $J = 4t^2/U$.
However, when $t \gg J$ ($U \to \infty$ limit), it is the motion of a hole that is responsible for the magnetism upon hole doping of such a Mott insulator.
Therefore, $SU(N)$ generalizations of the Nagaoka and counter-Nagaoka theorems are needed. 
In Refs. \cite{katsura2013nagaoka, 15puzzle2018exact}, it is shown that, with the ``unusual'' sign of the hopping matrix element, $t<0,$ a single hole motion in such a $U=\infty$ $SU(N)$ Hubbard model leads to a fully flavor-polarized ground state. 
However, less is understood for the same problem with the ``usual'' sign of hopping $t>0,$ again due to the frustration inherent in antiferromagnetism.

In this paper, we study the dynamics of a single hole doped at the $\frac 1 N$th filling of the $U=\infty$ $SU(N)$ Hubbard and $t$-$J$ models on certain solvable graphs.
We first consider such a problem on an $(N+1)$-site graph that satisfies the connectivity condition (as defined later), and show that the ground state is in the $SU(N)$ flavor-singlet sector (Sec. \ref{sec:a subgraph}).
Any other flavor configurations frustrate the hole motion.
From such an $(N+1)$-site subgraph, we construct a {\it subgraph tree}, on which the single hole problem in the $SU(N)$ $t$-$J$ model is exactly solvable (Secs. \ref{sec:a subgraph tree}-\ref{sec:tJ model}).
The ground state is a positive superposition of $SU(N)$ flavor-singlet covering states.
In Sec. \ref{sec:discussion}, we speculate on the possibility of exotic phases of matter in the presence of a dilute but finite hole concentration.

We note that the exact solvability of the single hole problem in a subgraph tree is due to the existence of an extensive number of local $SU(N)$ symmetries---in some sense, this is Hilbert space fragmentation \cite{rakovszky2020statistical, khemani2020localization, moudgalya2022quantum} from restricted hole motion.

\section{$SU(N)$ singlet in an $(N+1)$-site graph}
\label{sec:a subgraph}

We start by solving a single hole problem in the $U=\infty$ $SU(N)$ Hubbard model ($N \geq 2$) on an $(N+1)$-site graph that satisfies the connectivity condition (to be defined below). 
We assume that the hopping matrix elements are  positive but otherwise arbitrary $t_{ij} > 0$:
\begin{align}
\label{eq:Hamiltonian}
    \hat H &= -\sum_{\left <i,j \right >}\sum_{a=1}^N t_{ij} \left (c^{\dagger}_{i,a}c_{j,a} +{\rm H.c. }\right ) + \hat V(\{n_i\})
     + [U=\infty]. 
\end{align}
Here, $a=1, 2, ..., N$ is a flavor index of a fermion in the fundamental representation,  $i = 0, 1, 2, ..., N$ is a site index, and $\left < i,j \right >$ is an edge of the graph. 
$\hat V(\{n_i\}) $ describes  arbitrary on-site terms and density-density interactions  ($n_i \equiv \sum_{a=1}^Nc^{\dagger}_{i,a}c_{i,a}$):
\begin{align}
\label{eq:interactions}
    V(\{n_i\}) = \sum_i \epsilon_i n_i + \sum_{i,j} V_{ij}n_i n_j + \cdots.
\end{align}
The last $U=\infty$ term 
forbids any double occupancy. 

\smallskip

{\it Lemma}: The ground state of the Hamiltonian (\ref{eq:Hamiltonian}) on an $(N+1)$-site graph that satisfies the connectivity condition in the single hole sector is a unique $SU(N)$ flavor-singlet state.

\smallskip

In order to prove the {\it Lemma}, it is convenient to work in a particular many-body basis in a single hole sector.
In doing so, we restrict ourselves to a {\it flavor-balanced subspace}, where each flavor $a=1,2,...,N$ appears exactly once
\footnote{It can be shown straightforwardly that such a flavor-balanced subspace contains a state in every irreducible representation (irrep) that appears in $N$-direct products of the fundamental representation ${\bm N}$, ${\bm N}^N \equiv {\bm N} \times {\bm N} \times \cdots \times {\bm N}$.
Since any other states within each irrep can be reached by repeated applications of raising/lowering operators, it suffices to restrict ourselves to the flavor-balanced subspace.}.
For example, 
\begin{align}
\label{eq:initial MB basis}
    \left | \cdot, 1,2,...,N \right >  &\equiv 
    c^{\dagger}_{1,1}c^{\dagger}_{2,2} \cdots c^{\dagger}_{N,N} \vacuum \equiv \left | 0, 1,... , N \right> \nonumber \\
    &\equiv c_{0,0} c^{\dagger}_{0,0} c^{\dagger}_{1,1}c^{\dagger}_{2,2} \cdots c^{\dagger}_{N,N} \vacuum
\end{align}
is a flavor-balanced state, where $\vacuum$ is the vacuum state with no fermions and $0$ in the third expression denotes that the site $i=0$ is unoccupied.
This can be re-expressed as the final expression by creating a ghost fermion with flavor  $a=0$ at the hole site and annihilating it.
This is a useful notation that will be used throughout the paper. 
From this state, we form a complete orthonormal basis in a flavor-balanced subspace by applying a permutation of $(N+1)$ objects (a hole and $N$ fermions), $\sigma \in S_{N+1}$, where $S_{N+1}$ is the symmetric group of $(N+1)$ objects:
\begin{align}
\label{eq:basis}
    \left | \sigma \right > &\equiv  \left | \sigma(0),\sigma(1), ...,\sigma(N) \right >   \equiv  (-1)^{i} {\rm sgn}(\sigma) c^{\dagger}_{0,\sigma(0)}c^{\dagger}_{1,\sigma(1)} \times
    \nonumber \\ 
     &\ \ \ \ \ \ \cdots \times  c^{\dagger}_{i-1,\sigma(i-1)}c^{\dagger}_{i+1,\sigma(i+1)} \cdots c^{\dagger}_{N,\sigma(N)} \vacuum
    \nonumber \\
    &\equiv {\rm sgn}(\sigma)  c_{i,0 } c^{\dagger}_{0,\sigma(0)}c^{\dagger}_{1,\sigma(1)} \cdots c^{\dagger}_{N,\sigma(N)} \vacuum,
\end{align}
where we assumed that the $i$th site is occupied by a hole, i.e., $\sigma(i) = 0$ and again we introduced a ghost fermion with flavor $a=0$ in the last expression for  convenience.

Among the states in the flavor-balanced subspace are the completely flavor-antisymmetric, $SU(N)$ flavor-singlet (FS) states with the hole at site $i$,
\begin{align}
\label{eq:flavor-singlet}
    \left |i, \textrm{FS} \right > \equiv \frac 1 {\sqrt{N!}} \sum_{\sigma \in S_{N+1}, \sigma^{-1}(0) =i} \left | \sigma(0),\sigma(1), ...,\sigma(N) \right >.
\end{align}

{\it Connectivity condition}: An $(N+1)$-site graph is said to satisfy the connectivity condition if all the basis states in Eq. (\ref{eq:basis}) can be reached from one another by repeated applications of hopping operators in Eq. (\ref{eq:Hamiltonian}), $ \hat T_{ij} \equiv -t_{ij}\sum_{a=1}^N \left (c^{\dagger}_{i,a}c_{j,a} +{\rm H.c. }\right ).$ 
For example, Fig. \ref{fig:1} (a)-(c) are examples of graphs that satisfy the connectivity condition. 
In particular, in Fig. \ref{fig:1} (c), starting from the state $\left | 0,1,2,...,N\right >$, moving a hole around the triangular loop induces a transposition $(1\  2)$ and moving it around the largest, length $(N+1)$, loop induces the $N$-cycle $(1\ 2\ ...\ N)$. 
These two permutations, together with hopping operations, generate $S_{N+1}$.
More generally, Theorem 2 of Ref. \cite{15puzzle2018exact} provides the sufficient condition for the connectivity condition.

\begin{figure}[t]
    \centering
    \includegraphics[width = 0.5 \textwidth]{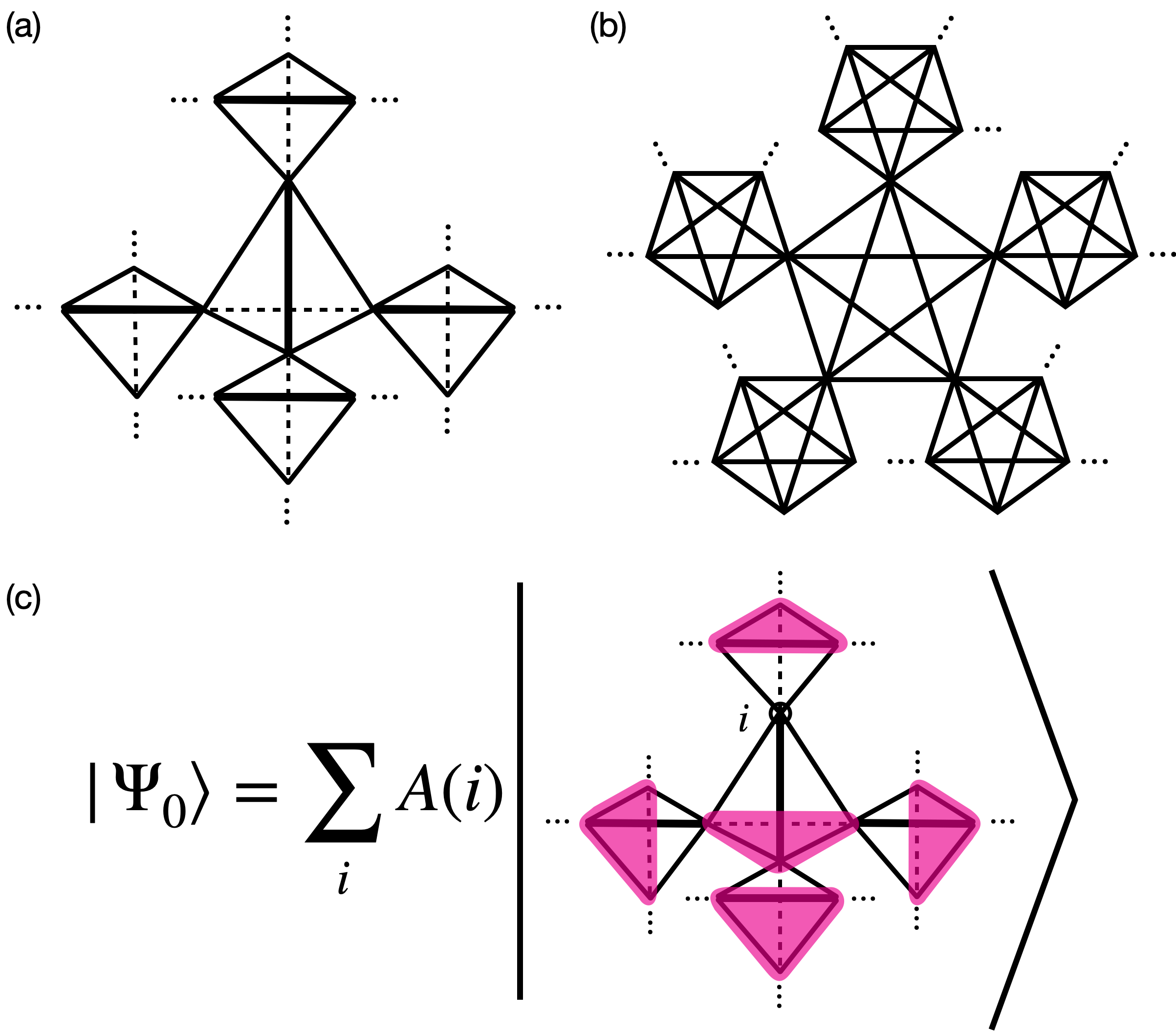}
    \caption{(a),(b) Examples of subgraph tree. In (a), sites are numbered in the way specified in the main text above Eq. (\ref{eq:initial MB basis for tree}). (c) The ground state in the single hole sector of the $U=\infty$ $SU(3)$ Hubbard [Theorem] and certain $t$-$J$ models [Corollary 2] is a positive  ($A(i)>0$) superposition of the $SU(3)$ flavor-singlet covering states. 
    }
    \label{fig:2}
\end{figure}

{\it Proof of the Lemma}: 
Any two basis states in the flavor-balanced subspace, $\left | \sigma \right> $ and $\left | \tau \right> $,  have a nonzero hopping matrix element only when they differ by one transposition involving a hole: $\sigma^{-1}(0) = \tau ^{-1}(a)$ and $\tau^{-1}(0) = \sigma ^{-1}(a)$ for some flavor $a$, and $\sigma^{-1}(k) = \tau^{-1}(k) $ for $k \neq 0,a$. 
Let $\sigma^{-1}(0) = i$ and $\tau^{-1}(0) = j$.
Any such nonzero off-diagonal matrix element is negative:
\begin{align}
\label{eq:matrix eleement for subgraph}
    \left< \sigma \right | \hat T_{ij} \left | \tau \right > = - t_{ij} <0.
\end{align}
Also,  the interaction term $V(\{n_i \})$ only contributes to diagonal matrix elements.
Therefore, the Perron-Frobenius theorem ensures that there exists a {\it unique} ground state $\left | \Psi_0 \right > $ which is a positive superposition of all the basis states ($A_{\sigma} > 0$):
\begin{align}
\label{eq:PF ground state Lemma}
    \left | \Psi_0 \right >  = \sum_{\sigma \in S_{N+1}} A_{\sigma} \left | \sigma \right >.
\end{align}
Since this state has a nonzero overlap with a flavor-singlet state $\left | i, \textrm{FS} \right >$, 
it must be a flavor-singlet state [if it were instead a superposition of multiple irreps of $SU(N)$, then it is possible to construct degenerate ground states, in contradiction to the uniqueness of the ground state]. 
Therefore, it is possible to rewrite Eq. (\ref{eq:PF ground state Lemma}) as a positive superposition ($A(i) >0$) of $\left | i, \textrm{FS} \right >$:
\begin{align}
\label{eq:Lemma}
    \left | \Psi_0 \right > = \sum_{i} A(i) \left | i, \text{FS} \right >.
\end{align}
See Fig. 1 (d) for the illustration of such a state. 
$\square$

\section{$SU(N)$ flavor-singlets in a subgraph tree}
\label{sec:a subgraph tree}
It is now straightforward to generalize the previous result to a ``subgraph tree" constructed as follows.
Starting from an $(N+1)$-site subgraph that satisfies the connectivity condition, we attach other $(N+1)$-site subgraphs to some (or all) of the vertices of the initial subgraph, in such a way that it does not create any other cycles (loops of length $l \geq 3$ in which only the first and the last vertices are equal) than those contained within each subgraph. 
This generates depth $1$ tree of $(N+1)$-site subgraph.
Continuing this $n$ times will generate a depth $n$ subgraph tree, which has the property that all the  cycles of the graph are contained within each subgraph. 
Let $N_{\rm SG}$ be the number of subgraphs constituting such a subgraph tree. 
The number of sites in such a graph is $N  N_{\rm SG } +1$.
Figures 2 (a-b) are examples of such graphs. 
We will consider the Hamiltonian (\ref{eq:Hamiltonian}) on such a graph in the presence of a single hole.

The advantage of such a subgraph tree is that there is an $SU(N)$ symmetry associated with each subgraph as can be seen as follows \footnote{Ref. \cite{kim2023RVB} deals with a special case $N=2$.}.
First, a many-body basis can be constructed by locating the site of the hole $i,$ and then specifying the flavor configuration on the rest of the sites. 
Once the hole location is specified, it is easy to see that there is a unique $N$-mer covering of the lattice [see Fig. 2 (c) for the illustration of such a covering].
In any step in which the hole hops to a neighboring site, one $N$-mer is moved, but in such a way that it remains inside the initial $(N+1)$-site subgraph in which it was contained.  
Thus, we can label the $N$-mers uniquely by a subgraph index $s=1,...,N_{\rm SG}$, and the total flavor $SU(N)$ symmetry is preserved for $N$ fermions contained in each $s$ $N$-mer.
That is, there exists $SU(N)^{N_{\rm SG}} = SU(N) \otimes SU(N) \otimes \cdots \otimes SU(N)$ symmetry.

Thanks to such $SU(N)^{N_{\rm SG}}$ symmetry, it is enough to consider a subspace that is flavor-balanced {\it in each} $s$ $N$-mer.
Any other states in the Hilbert space can be reached by repeated applications of raising and lowering operators on each $s$ $N$-mer. (See Appendix \ref{sec:raising and lowering} for the expression of those raising/lowering operators.)
We now construct a many-body basis restricted in such a flavor-balanced subspace analogously to Eqs. (\ref{eq:initial MB basis}) and (\ref{eq:basis}).
We first occupy a hole at a particular location (call it $i=0$), which will define a unique $N$-mer covering as discussed above.
For each $s$ $N$-mer, we label the sites contained in it by $i = (s-1)  N +1 ,\ (s-1)  N +2 , \cdots,\  s N $  (See Fig. 2. (a) for the illustration of such a site numbering scheme along with subgraph indices $s$) and occupy it with fermions with flavors $a = 1,\cdots,N,$ respectively. 
This defines one basis state 
\begin{align}
    \label{eq:initial MB basis for tree}
    &\left | 0, \left (1,... , N \right ), \left(  1,...,N \right ),  ... , \left(  1,...,N \right ) \right> \nonumber \\ \equiv& c_{0,0}c_{0,0}^{\dagger}
    \left( c^{\dagger}_{1,1}c^{\dagger}_{2,2} \cdots c^{\dagger}_{N,N} \right)
    \left( c^{\dagger}_{N+1,1} \cdots c^{\dagger}_{2N,N} \right) \nonumber \\ &
    \times \cdots \times \left( c^{\dagger}_{N\cdot(N_{\rm SG}-1)+1,1} \cdots c^{\dagger}_{N\cdot N_{\rm SG},N} \right) \vacuum,
\end{align}
where again, the ghost flavor index $a=0$ is introduced for convenience in $c_{0,0}$.
Using the fact that fermions in different $N$-mers do not exchange one another due to the restricted dynamics of a hole, we might as well treat them as distinguishable and rename a flavor index $a$ in $s$ $N$-mer to be $(s-1)N + a$. 
Hence, the basis state (\ref{eq:initial MB basis for tree}) can be denoted by
\begin{align}
    \label{eq:MB basis renaming}
    &\left | 0, 1,... , N ,N+1  ... ,  ... , N\cdot N_{\rm SG} \right> \nonumber \\ \equiv& c_{0,0}c_{0,0}^{\dagger}
    \left( c^{\dagger}_{1,1}c^{\dagger}_{2,2} \cdots c^{\dagger}_{N,N} \right)
    \left( c^{\dagger}_{N+1,N+1} \cdots c^{\dagger}_{2N,2N} \right) \nonumber \\ &
    \times \cdots \times \left( c^{\dagger}_{N(N_{\rm SG}-1)+1,N(N_{\rm SG}-1)+1} \cdots c^{\dagger}_{N N_{\rm SG},N N_{\rm SG}} \right) \vacuum.
\end{align}
From this state, any other basis state that is flavor-balanced for each $s$ $N$-mer can be reached by repeated applications of hopping operators $\hat T_{ij}$.
There are $(N N_{\rm SG} +1) (N!)^{N_{\rm SG}}$ different such (orthonormal) basis states.
Each such basis state has a permutation operator $\sigma \in S_{NN_{\rm SG}+1}$ associated with it defined as a relative flavor configuration from the initial one in Eq. (\ref{eq:MB basis renaming}): 
 if site $i$ is occupied by the flavor $a$, then $\sigma(i) \equiv a$.
(We emphasize that flavor indices are renamed to have values $a=0,1,...,NN_{\rm SG}$).
Let $P$ be the collection of all such permutations $\sigma$.
We define the basis states $\{ \left |\sigma \right > :\  \sigma \in P \}$  with a particular sign structure analogous to Eq. (\ref{eq:basis}):
\begin{align}
    \label{eq:MB basis for tree}
    \left | \sigma \right > \equiv &\left | \sigma(0), ... , \sigma(N N_{\rm SG}) \right>  \equiv  (-1)^{i} {\rm sgn}(\sigma) c^{\dagger}_{0,\sigma(0)}c^{\dagger}_{1,\sigma(1)}
    \nonumber \\
    & \times \cdots \times   c^{\dagger}_{i-1,\sigma(i-1)}c^{\dagger}_{i+1,\sigma(i+1)} \cdots c^{\dagger}_{NN_{\rm SG},\sigma(NN_{\rm SG})} \vacuum
    \nonumber \\
    =& {\rm sgn}(\sigma)  c_{i,0 } c^{\dagger}_{0,\sigma(0)}c^{\dagger}_{1,\sigma(1)} \cdots c^{\dagger}_{NN_{\rm SG},\sigma(NN_{\rm SG})} \vacuum,
\end{align}
where we again assumed that the $i$th site is occupied by a hole, i.e., $\sigma(i) = 0$.
The sign structure again allows us to write the $SU(N)$ flavor-singlet covering (FSC) state, the state with an $SU(N)$ flavor-singlet on every $N$-mer, as a uniform superposition of the basis states that have their hole at site $i$:
\begin{align}
\label{eq:flavor-singlet covering}
    &\left |i, \textrm{FSC} \right >   \equiv \left | i, \textrm{FS}_1, \cdots, \textrm{FS}_{N_{\rm SG}} \right >
    \nonumber \\ 
    &= 
    \frac 1 {\sqrt{(N!)^{N_{\rm SG}}}} \sum_{\sigma \in P \atop \sigma(i) =0} \left | \sigma(0),\sigma(1), ...,\sigma(N N_{\rm SG}) \right >.
\end{align}
The following Theorem is the main result of our paper.

\smallskip
{\it Theorem}: The ground state of the Hamiltonian (\ref{eq:Hamiltonian}) on a ``subgraph tree" in the single hole sector is unique and is a positive ($A(i)>0$) superposition of the $SU(N)$ flavor-singlet covering (FSC) states \footnote{We note that when  hopping matrix elements $t_{ij}$ and on-site potentials $\epsilon_i$ satisfy certain relations, the Theorem reproduces the result of the Brandt-Giesekus model \cite{brandtGiesekus1992,mielke1992exact,strack1993exact,tasaki1993exactRVB}}:
\begin{align}
\label{eq:Theorem}
    \left |\Psi_0 \right > = \sum_{i} A(i) &\left |i, \textrm{FSC} \right >. 
\end{align}
(See Fig. 2 (c) for an illustration of this state)

\smallskip

{\it Proof of Theorem}: It is straightforward to show that any nonzero off-diagonal element of the Hamiltonian matrix is negative, $\left< \sigma \right | \hat T_{ij} \left | \tau \right > = - t_{ij} <0$, as in Eq. (\ref{eq:matrix eleement for subgraph}). 
 Also, since any basis state $\left | \sigma\right >$ can be reached from one another by repeated applications of $\hat T_{ij}$, one concludes from the Perron-Frobenius theorem that the ground state is unique and is a positive ($A_{\sigma}>0$) superposition of all the basis states:
 \begin{align}
\label{eq:PF ground state}
    \left | \Psi_0 \right >  = \sum_{\sigma \in P} A_{\sigma} \left | \sigma \right >.
\end{align}
This has a positive overlap with a flavor-singlet covering state $\left| i ,\textrm{FSC} \right >$, and hence $N$ fermions in every $N$-mer must be a flavor-singlet.
Hence, $\left | \Psi_0 \right > $ can be rewritten as a superposition of flavor-singlet covering states as in Eq. (\ref{eq:Theorem}). $\square$

In Fig. \ref{fig:3} of Appendix \ref{sec:ED}, we show the result of the finite size exact diagonalization study on the $SU(3)$ $U=\infty$ Hubbard model on the four-tetrahedron geometry. The result agrees with the Theorem.

\section{$SU(N)$ $t$-$J$ model} 
\label{sec:tJ model}
Now we generalize the previous results to the $SU(N)$ $t\mbox{-}J$ model.
In the presence of a finite but large $U$ ($\gg t$) term, $\frac U 2 \sum_i \hat n_i(\hat n_i -1 )$, 
one can obtain the $SU(N)$ $t\mbox{-}J$ model from the $SU(N)$ Hubbard model by projecting out the states with multiply occupied sites \cite{macdonaldGirvinYoshioka1988, hirsch1985attractive, oitmaa2006series}:
\begin{align}
    \label{eq:t-J model}
    \hat H_{t\mbox{-}J} &= \hat H + \sum_{\left < i,j\right >}J_{ij} \left( \hat {\boldsymbol{\lambda}}_i \cdot \hat {\boldsymbol{\lambda}}_j - \frac{N-1}{2N} \hat n_i \hat n_j \right)
    \nonumber \\
    &- \sum_{\left < i,j,k \right >}\sum_{1\leq a < b \leq N}
    K_{ijk} \hat \Delta_{jk}^{ab \dagger}\hat \Delta_{ij}^{ab} + O \left( \frac{t^3}{U^2} \right)
    \nonumber \\
    &\equiv \hat H + \sum_{\left < i,j\right>}\hat J_{ij} + \sum_{\left < i,j,k\right>}\hat K_{ijk}+ O \left( \frac{t^3}{U^2} \right)
    .
\end{align} 
Here $\hat H$ is the Hamiltonian for the $U=\infty$ Hubbard model (\ref{eq:Hamiltonian}), $J_{ij} = 4t_{ij}^2/U$ and $K_{ijk} = 2t_{ij}t_{jk}/U$, $\left < i,j,k \right >$ denotes the triplet of sites such that $j$ is a nearest neighbor to $i$ and $k$, and $\hat \Delta_{ij}^{ab} \equiv \frac 1 {\sqrt 2 } (c_{i,a}c_{j,b}-c_{i,b}c_{j,a})$ is the annihilation operator of a flavor-antisymmetric state on a bond $\left < i,j \right >$.
$\hat {\boldsymbol \lambda}_{i} = (\hat \lambda_{i}^{(1)}, ..., \hat \lambda_{i}^{(N^2-1)})$ denotes $(N^2-1)$ generators of the $SU(N)$ group at site $i$ with the normalization ${\rm Tr}(\lambda_i^{(r)} \lambda_j^{(r')}) = \frac 1 2 \delta_{r,r'} \delta_{i,j}$ \cite{georgi2000lie}.
The Heisenberg operator can be rewritten in terms of a flavor-permutation operator $\hat P_{ij}$ as
$J_{ij}  ( \hat {\boldsymbol \lambda}_{i} \cdot \hat {\boldsymbol \lambda}_{j} - \frac {N-1} {2N} \hat n_i \hat n_j) = \frac 1 2 J_{ij}(\hat P_{ij}-  \mathbb{\hat 1}) \hat n_i \hat n_j$.
In the last line, we defined $\hat J_{ij} \equiv J_{ij}(\hat {\boldsymbol{\lambda}}_i \cdot \hat {\boldsymbol{\lambda}}_j - \frac{N-1}{2N} \hat n_i \hat n_j)$ and $\hat K_{ijk} \equiv K_{ijk} \sum_{1 \leq a < b \leq N} \hat \Delta_{jk}^{ab \dagger}\hat \Delta_{ij}^{ab} $.
The following two corollaries generalize the Lemma and the Theorem to certain $t$-$J$ models.

\smallskip

{\it Corollary 1}:
Let us define $\hat H_{t\mbox{-}J}$ on an $(N+1)$-site graph that satisfies the connectivity condition. 
$J_{ij}\geq 0$ and $K_{ijk}\geq 0$ do not have to be related to one another and can be arbitrary independent parameters.
Then, the ground state of $\hat H_{t\mbox{-}J}$ in the presence of a single hole is unique and is a positive superposition of flavor-singlet states (\ref{eq:Lemma}) as in the Lemma.

\smallskip

{\it Corollary 2}:
For $\hat H_{t\mbox{-}J}$ defined on a subgraph tree, 
let $J_{ij} = J_s \geq 0$ be uniform within each subgraph and connect any two sites within it. 
Also, let $K_{ijk}\geq 0$ terms act only on three sites $\left <i,j,k \right >$ fully contained within a subgraph.
Again, $J_{ij}$ and $K_{ijk}$  can be independent parameters, except for the above constraints.
Then, the ground state of $\hat H_{t\mbox{-}J}$ in the presence of a single hole is unique and is a positive superposition of flavor-singlet covering states (\ref{eq:Theorem}) as in the Theorem \footnote{The results of Ref. \cite{kim2023RVB} on the $SU(2)$ $t$-$J$ model on a triangular cactus is the $N=2$ case of  Corollary 2.}.

\smallskip

{\it Proof of Corollary 1}: 
For a single hole problem in an $(N+1)$-site graph, the total $SU(N)$ symmetry is intact even in the presence of $\hat J$ and $\hat K$ terms, and one can work in the flavor-balanced basis (\ref{eq:basis}). 
Again, it is then sufficient to show that all the nonzero off-diagonal matrix elements are negative.
In particular, for $\sigma \neq \tau$, $\left < \sigma \right | \hat J_{ij} \left | \tau \right >$ is nonzero only when $\sigma$ and $\tau$  differ by one transposition between occupied sites: $\sigma(i)= \tau(j) \neq 0$, $\sigma(j)= \tau(i)\neq 0$, and $\sigma(k) =\tau(k)$ for $k\neq i,j$.
In such a case, one obtains 
\begin{align}
    \left  <\sigma \right | \hat J_{ij} \left | \tau \right > = -J_{ij}/2 <0.
\end{align}
Similarly, any nonzero off-diagonal element of 
$\hat K_{ijk}$ is negative
\begin{align}
\label{eq:K term}
\left < \sigma \right | \hat K_{ijk} \left | \tau \right > = -K_{ijk}/2 <0.
\end{align}
This completes the proof.
$\square$

{\it Proof of Corollary 2}: 
For a subgraph tree, consider first the case when $\hat J=0$.
When $K_{ijk}$ are nonzero only for triplets of sites $\left < i,j,k \right >$ fully contained in a subgraph, $SU(N)^{N_{\rm SG}}$ symmetry is intact.
Thus, one can still work in the flavor-balanced basis (\ref{eq:MB basis for tree}) and the same proof as in the Theorem can be applied.

When $\hat J \neq 0$,  the $SU(N)^{N_{\rm SG}}$ symmetry is lost.
However, for the special case where $J_{ij}= J_s$ is uniform within each subgraph and connects any two sites within it, one can rewrite the Heisenberg term as [density-density interactions in $\hat J$ can be absorbed in $\hat V $ term in Eq. (\ref{eq:Hamiltonian})]:
\begin{align}
   \sum_{\left < i,j\right >}J_{ij}\hat {\boldsymbol{\lambda}}_i \cdot \hat {\boldsymbol{\lambda}}_j = \sum_{s=1}^{N_{\rm SG}} \frac{J_s}{2}\left [  \left ( \sum_{i=1}^{N+1} \hat {\boldsymbol \lambda}_{(s,i)} \right )^2 -  \sum_{i=1}^{N+1}  \hat {\boldsymbol \lambda}_{(s,i)} ^2  \right ].
\end{align}
Here $(s,i)$ denotes the site $i=1,2,...,N+1$ in a subgraph $s$. 
This Heisenberg operator takes the lowest possible eigenvalue for the flavor-singlet covering states (\ref{eq:flavor-singlet covering}):
\begin{align}
    \sum_{\left < k,l\right >}J_{kl}\hat {\boldsymbol{\lambda}}_k \cdot \hat {\boldsymbol{\lambda}}_l\left |i, \textrm{FSC} \right > = -\frac {N^2-1}{4}\sum_{s=1}^{N_{\rm SG}}J_s\left |i, \textrm{FSC} \right >.
\end{align}
This means that the ground state of $\hat H_{t\mbox{-}J}$ is still in the flavor-singlet covering subspace spanned by states (\ref{eq:flavor-singlet covering}) and
is  of the form Eq. (\ref{eq:Theorem}) with positive $A(i) >0$.
$\square$

\section{Discussion}
\label{sec:discussion}
Our result demonstrates the fundamental importance of the sign of the hopping matrix elements $t_{ij}$ on a kinetic magnetism of the Hubbard model, which in turn, manifests as a particle-hole asymmetry in the magnetic phase diagram. 
More precisely, in the usual $SU(2)$ Hubbard model, the particle-hole transformation $c_{i,\sigma} \rightarrow c_{i,\sigma}^{\dagger}$, with $\sigma = \uparrow,\downarrow$, maps the single doublon problem to the single hole problem with the opposite sign of $t_{ij}$ \cite{arovas2021hubbard}.
This implies that for a bipartite lattice---where the sign of $t_{ij}$ can be changed by a gauge transformation---the phase diagram is particle-hole symmetric around half-filling.
On the other hand, for a non-bipartite lattice the phase diagram exhibits a particle-hole asymmetry.
For example, the single hole dynamics in the triangular lattice $U=\infty $ Hubbard model leads to a $120^{\circ}$ antiferromagnetic ordering \cite{haerter2005kineticAF, sposetti2014kineticAF} whereas the single doublon problem satisfies the Nagaoka theorem and leads to a fully polarized ferromagnet (except for one singlet for a doublon).
Performing such a particle-hole transformation to the $SU(N)$ Hubbard model, one maps a single hole problem at $1/N$ filling to a single $N$-on ($N$ fermions at a site) problem at $(N-1)/N$ filling with the opposite sign of $t_{ij}$.
Since $(N-1)$ fermions at the same site must be completely flavor-antisymmetric, such $N-1$ electrons form a complex conjugate representation  $\bar {\bm N}$  of the fundamental representation.
Hence we see that with the usual sign of the hopping $t_{ij}>0$, while the Nagaoka ferromagnetic state appears for a single fermion doping of the $({N-1})/ N$ filled Mott insulator, a single hole doping at $1/N$ filling generically induces antiferromagnetism.

We note that in a more realistic non-bipartite lattice (e.g. a triangular or pyrochlore lattice), it is likely that the hopping operators $\hat T$ and exchange interactions $\hat J$ (or singlet hopping terms $\hat K$) favor different local magnetic correlations.
In such a case, the hole can only delocalize in a finite number of sites, leading to the formation of an RVB polaron.

Going from such a single RVB/RFS polaron problem to a multi-polaron (or multi-hole) problem requires yet another technical development, but we can speculate on possible outcomes (apart from a trivial phase separation scenario).
First, it is possible to have a broken-$SU(N)$-symmetry phase with a long-range flavor-antiferromagnetic order when flavor-singlets are supported over a sufficiently long distance \cite{liangDoucotAnderson1988, haerter2005kineticAF, sposetti2014kineticAF, zhu2022doped}. 
When $SU(N)$ flavor-singlets are supported on sufficiently short distances, one can have a flavor-disordered phase with topological order \cite{rokhsarKivelson1988QDM,moessnerSondhi2001,leeOhHanKatsura, Z3rydberg,SU(3)spinLiquid}.
The flavor-disordered state with a broken translation symmetry corresponds to various topologically ordered  crystalline phases \cite{jiang2017holonWC,kim2021quantum}.
Finally, it is possible to have various exotic liquid phases with a topological character such as a $\mathbb{Z}_N$ topologically ordered  Fermi liquid (FL* phase) \cite{Senthil_FL*} or high-temperature superconductivity.

\section*{Acknowledgement}
K-S.K. would like to acknowledge the hospitality of the Massachusetts Institute of Technology, where the majority of this work was done.
K-S.K. appreciates Samuel Alipour-fard, Zhaoyu Han, and Pavel Nosov for helpful discussions.
K-S.K. also thanks Aidan Reddy for valuable discussions on the exact diagonalization method.
H.K. would like to thank Tadahiro Miyao for useful comments on the manuscript. 
K-S.K. was supported by the Department of Energy, Office of Basic Energy Sciences, Division of Materials Sciences and Engineering, under contract DE-AC02-76SF00515.
H.K. was supported by JSPS KAKENHI Grants No. JP18K03445, No. JP23H01093, No. 23H01086, and MEXT KAKENHI Grant-in-Aid for Transformative Research Areas A “Extreme Universe” (KAKENHI Grant No. JP21H05191).

\appendix

\section{Exact diagonalization result}
\label{sec:ED}
\begin{figure}[h]
    \centering
\includegraphics[width = 0.49 \textwidth]{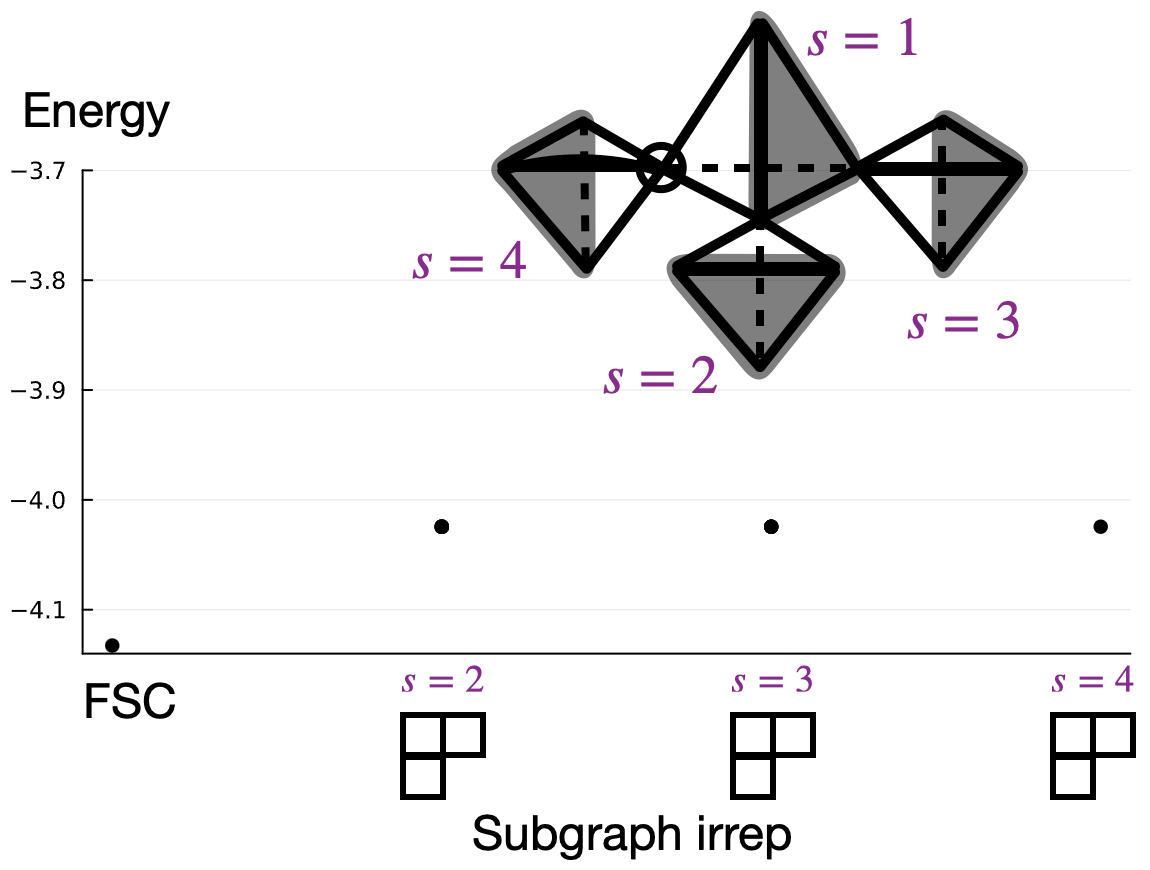}
    \caption{Ground state and several first excited states of $SU(3)$ $U=\infty$ Hubbard model on the four-tetrahedron geometry.
The horizontal axis denotes the Young diagram corresponding to the irrep of the $SU(3)$ group of the trimer (shown in gray) in the specified subgraph $s$. 
    The three first excited states shown are the ones that have adjoint representation  
    in $s=2,3,4$ trimer, respectively; trimers in the subgraphs that are not specified are  $SU(3)$ singlets. 
    FSC is the sector which has $SU(3)$ singlet on every trimer. 
    The ground state is of the form Eq. (13)
    and has the same energy as the non-interacting problem with the same hopping matrix elements, $E_{\rm
 GS} = -4.1326383.$
    }
    \label{fig:3}
\end{figure}

In order to demonstrate the result of our main Theorem, we performed exact diagonalization calculation on the $SU(3)$ $U=\infty$ Hubbard model in the single-hole sector on a four-tetrahedron geometry as shown in Fig. \ref{fig:3} ($13$-sites; shown in the inset).
Hopping matrix elements are set to $t_{ij}= 1$ for all bonds.
The ground state is the positive superposition of $SU(3)$ flavor-singlet covering states, consistent with the result of the Theorem.

\section{Raising and lowering operators}
\label{sec:raising and lowering}

\begin{figure}[t]
    \centering
\includegraphics[width = 0.45 \textwidth]{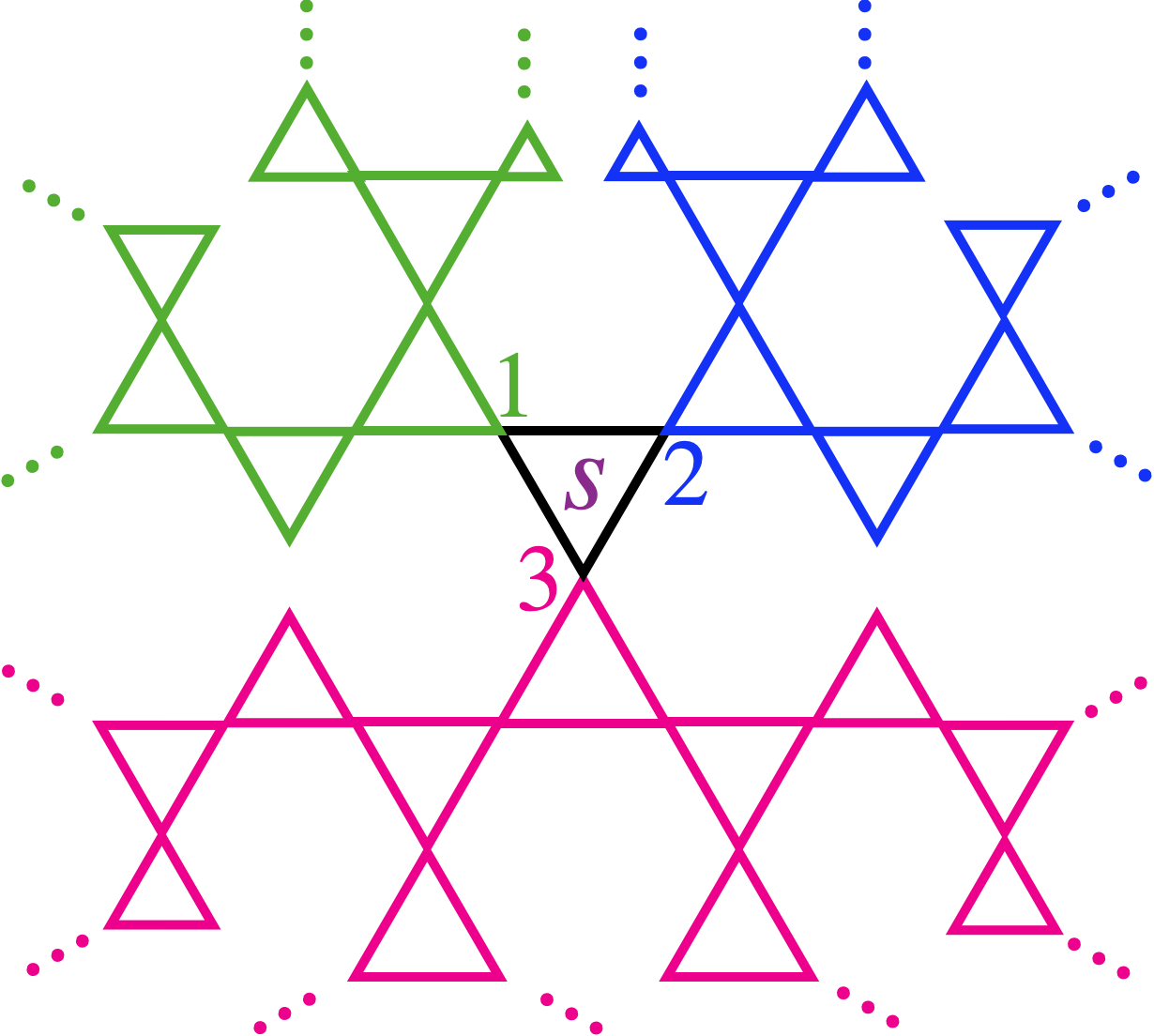}
    \caption{One can classify sites according to the index of a branch stemming from $s$th subgraph. Green, blue, and magenta sites are in the 1st, 2nd, and 3rd branches of the $s$ subgraph. The set of all sites contained in each branch can be denoted by $\mathcal 
V_1$, $\mathcal V_2$, and $\mathcal V_3$. $\mathcal V_1 \cup\mathcal  V_2 \cup \mathcal V_3$ is the set of all sites in the graph.}
    \label{fig:sup}
\end{figure}

The local $SU(N)$ symmetry on each subgraph of the $U=\infty$ $SU(N)$ Hubbard model in the single-hole sector allows one to define raising and lowering operators on each $s$ $N$-mer.
Such operators for the $SU(2)$ case in terms of fermionic operators can be written as follows:
\begin{align}
    &\hat {\mathscr S}^{\pm}_{s} = \sum_{i \in \mathcal V_1} \hat h_i \left( \prod_{j \notin \{i,2,3\} } \hat n_j \right) \hat S^{\pm}_{(23)}+ \nonumber \\
    &\sum_{i \in \mathcal V_2} \hat h_i \left( \prod_{j \notin \{i,1,3\} } \hat n_j \right) \hat S^{\pm}_{(31)}+\sum_{i \in \mathcal V_3} \hat h_i \left( \prod_{j \notin \{i,1,2\} } \hat n_j \right) \hat S^{\pm}_{(12)}.
\end{align}
Here $\hat h_i = 1- \hat n_i$ is the hole number operator at site $i$ and $\hat S^{\pm}_{(ij)}$ is the raising/lowering operator of the total spin of two fermions at sites $i$ and $j.$ 
$\mathcal V_{1,2,3}$ are the set of sites defined in Fig. \ref{fig:sup}.
The operator $\hat {\mathscr S}_{s}^{\pm}$ is defined for every $s=1,...,N_{\rm SG}$.
{\it In the single hole sector}, it is straightforward to show that they commute with the hopping operators and number operators, and hence with the Hamiltonian
$\hat H$ (\ref{eq:Hamiltonian}) of the main text:
\begin{align}
&\hat P_{\hat N_h =1}[\hat {\mathscr S}_s^{\pm}, \hat T_{ij}]\hat P_{\hat N_h =1}= 0, \nonumber \\
&\hat P_{\hat N_h =1}[\hat {\mathscr S}_s^{\pm}, \hat n_i]\hat P_{\hat N_h =1}= 0,
\nonumber \\
&\hat P_{\hat N_h =1}[\hat {\mathscr S}_{s}^{\pm}, \hat H]\hat P_{\hat N_h =1}=0\ \ \ \forall s.
\end{align}
Here, $\hat P_{\hat N_h =1}$ is the projection to the single hole sector and  $\hat T_{ij}$ is the electron hopping operator defined in the main text: $ \hat T_{ij} \equiv -t_{ij}\sum_{a=1}^N \left (c^{\dagger}_{i,a}c_{j,a} +{\rm H.c. }\right ).$ 
Note that ``local'' raising/lowering operators $\hat {\mathscr S}_s^{\pm}$ become non-local in terms of fermion operators.

\section{Derivation of Eq. (19)}
\label{sec:appendix C}

Here, we show that the smallest eigenvalue of the following operator defined for each subgraph is  $-\frac {N^2-1}{2}$ when the number of electrons is $N$ or $N+1$:
\begin{align}
     \left ( \sum_{i=1}^{N+1} \hat {\boldsymbol \lambda}_{i} \right )^2 -  \sum_{i=1}^{N+1}\left (  \hat {\boldsymbol \lambda}_{i} \right )^2. 
\end{align}
$\hat  {\boldsymbol \lambda}_i \equiv \sum_{a,b = 1}^N c^{\dagger}_{i,a} [{\boldsymbol \lambda}_i]_{a,b} c_{i,b}$ are $SU(N)$ generators in the fundamental representation in terms of fermion operators with normalization ${\rm Tr}(\lambda_i^{(r)} \lambda_j^{(r')}) = \frac 1 2 \delta_{r,r'} \delta_{i,j}$.
First, when the number of fermions is $N+1,$ every site of the subgraph is occupied, so $\left ( \hat {\boldsymbol \lambda}_i\right )^2 = \frac{N^2-1}{2N}\cdot \mathbb{\hat 1}.$ 
Also, the Casimir operator has the smallest eigenvalue when $N$ of the fermions form a singlet [i.e., when $N+1$ fermions form a fundamental representation of $\sum_{i=1}^{N+1} \hat {\boldsymbol \lambda}_i$], so that $\left (\sum_{i=1}^{N+1} \hat {\boldsymbol \lambda}_i\right )^2 = \frac{N^2-1}{2N}\cdot \mathbb{\hat 1}.$ 
In such a case, $\left ( \sum_{i=1}^{N+1} \hat {\boldsymbol \lambda}_{i} \right )^2 -  \sum_{i=1}^{N+1}\left (  \hat {\boldsymbol \lambda}_{i} \right )^2 = -\frac{N^2-1}{2}\cdot \mathbb{\hat 1}$.

On the other hand, when the number of fermions is $N$, $\left ( \sum_{i=1}^{N+1} \hat {\boldsymbol \lambda}_{i} \right )^2$ has the smallest eigenvalue, $0,$ when they form an $SU(N)$ singlet. Again, this implies $\left ( \sum_{i=1}^{N+1} \hat {\boldsymbol \lambda}_{i} \right )^2 -  \sum_{i=1}^{N+1}\left (  \hat {\boldsymbol \lambda}_{i} \right )^2 = -\frac{N^2-1}{2}\cdot \mathbb{\hat 1}$.

\bibliography{ref.bib}

\end{document}